\documentclass[aps,prl,superscriptaddress,twocolumn]{revtex4}
\usepackage{graphicx}
\usepackage{epstopdf}
\usepackage{bm}
\usepackage{amssymb}
\usepackage{color}

\begin{document}

\title{Helimagnons in a chiral ground state of the pyrochlore antiferromagnets}
\author{Eunsong Choi}
\affiliation{Department of Physics, University of Wisconsin, Madison, Wisconsin 53706, USA}

\author{Gia-Wei Chern}
\affiliation{Department of Physics, University of Wisconsin, Madison, Wisconsin 53706, USA}
\affiliation{Theoretical Division, Los Alamos National Laboratory, Los Alamos, New Mexico 87545, USA}

\author{Natalia B. Perkins}
\affiliation{Department of Physics, University of Wisconsin, Madison, Wisconsin 53706, USA}

\date{\today}

\begin{abstract}
The Goldstone mode in a helical magnetic phase, also known as the helimagnon,
is a propagating mode with a highly anisotropic dispersion relation. Here we study theoretically
the helimagnon excitations in a complex chiral ground state of pyrochlore antiferromagnets
such as spinel CdCr$_2$O$_4$ and itinerant magnet YMn$_2$. We show that the effective
theory of the soft modes in the helical state possesses a symmetry similar to that of
smectic liquid crystals. We compute the low-energy spin-wave spectrum based on a microscopic
spin Hamiltonian and find a dispersion relation characteristic of the helimagnons.
By performing dynamics simulations with realistic model parameters,
we also obtain an overall agreement between experiment and the numerical spin-wave spectrum.
Our work thus also clarifies the mechanisms that relive the magnetic frustration in these compounds.
\end{abstract}

\maketitle

{\em Introduction.}
The rich phenomenology associated with helical spin ordering has attracted considerable attention
recently. Magnetic spirals have been shown to play a crucial role in inducing the spontaneous
electric polarization in multiferroic materials~\cite{mostovoy06}. The soft magnetic fluctuations in a helical
spin-density-wave state of itinerant magnet MnSi also holds the key to its non-Fermi liquid behavior~\cite{pfleiderer04}.
Contrary to most well known long-range spin orders such as ferromagnetism and antiferromagnetism,
spin-wave excitations in a helical magnetic order is described by a theory similar to
the elasticity of smectic liquid crystals~\cite{deGennes93,chaikin}.
In particular, despite the broken symmetry of the helical
order is described by a O(3) order parameter, there only exists a single Goldstone mode similar to the
smectic-like phonon mode which emerges on scales larger than the helical pitch~\cite{belitz06,radzihovsky11}.
This low-energy gapless excitations, dubbed helimagnons in Ref.~\cite{belitz06}, exhibit a highly anisotropic dispersion relation.

\begin{figure}
\centering{}\includegraphics[width=1\columnwidth]{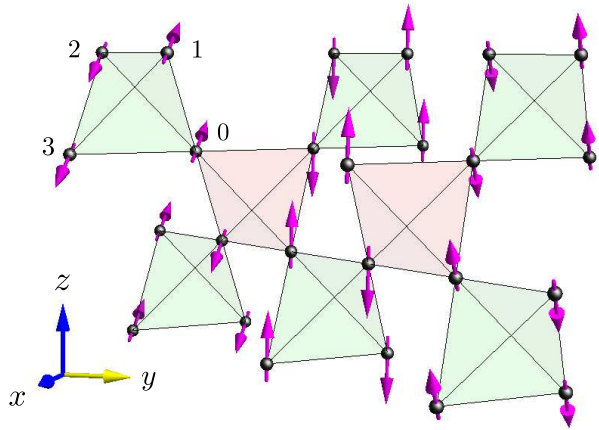}
\caption{(Color online) Coplanar helical order with spins lying in the $ac$ plane and rotating about the $b$ axis.
The magnetic order corresponds to the solution $\hat\mathbf n_1 = (\cos\theta(y), 0, \sin\theta(y))$,
with $\theta(y) = Q y + \mbox{const}$ with $Q=2\pi\delta$. It produces a Bragg peak $\mathbf Q = 2\pi(0, \delta, 1)$ in
neutron scattering, consistent with the experimental measurement on CdCr$_2$O$_4$.
The exchange anisotropies are $K_{[0,1,\pm1]} = K'_{[\pm1, 0, 1]} = -2K_u$, and $K_{[\pm1, 0, 1]} = K'_{[0,1,\pm1]} = K_{[1,\pm 1,0]} = K'_{[1,\pm 1,0]} = K_u$, where $K$ and $K'$ denote exchange anisotropy on
NN bonds of the green and red tetrahedra, and the subscript indicates the bond direction.
\label{fig:spiral}}
\end{figure}

The recent discovery of a novel chiral spin structure in spinel CdCr$_2$O$_4$
has generated much interest both theoretically and experimentally~\cite{chern06,chung05,matsuda07,bhattacharjee11,
valdes08,kant09,kim11,kimura06}.
A very similar coplanar helical order has also been reported in the weak itinerant antiferromagnet YMn$_2$~\cite{ballou87,cywinski91}.
The magnetic ions in these compounds form a three-dimensional network of
corner-sharing tetrahedra (Fig.~\ref{fig:spiral}), known as the pyrochlore lattice.
The geometrically frustrated spin interactions in this lattice
gives rise to an extensively degenerate ground state at the classical level~\cite{moessner98}. As a result,
magnetic ordering at low temperatures is determined by the dominant residual perturbations in the system.

The helical magnetic order in CdCr$_2$O$_4$ is stabilized by a combined effect of
magnetoelastic coupling and relativistic spin-orbit interactions.
Despite a rather high Curie-Weiss temperature $|\Theta_{\rm CW}| \approx 70$ K,
the spiral magnetic order sets in only at $T_N = 7.8$ K, indicating a high degree of frustration.
The magnetic transition at $T_N$ is accompanied by a structural distortion which
lowers the crystal symmetry from cubic to tetragonal~\cite{chung05}.
Recent infrared absorption measurements indicated a broken lattice inversion symmetry below $T_N$~\cite{kant09},
consistent with the scenario predicted in Ref.~\cite{chern06} that the magnetic frustration is
relieved by the softening of a $\mathbf q = 0$ optical phonon with odd parity.
A long-range N\'eel order with ordering wavevector $\mathbf q = 2\pi(0,0,1)$ is stabilized
by the lattice distortion. Moreover, the lack of inversion symmetry endows the crystal structure
with a chirality. The collinear state is transformed into a magnetic spiral as the chirality
is transferred to the spins through spin-orbit interaction. The shifted ordering vector $\mathbf q = 2\pi(0,\delta,1)$
is consistent with the observed magnetic Bragg peaks.

In this paper we undertake a theoretical calculation of the spin-wave spectrum in the helical ground state
of pyrochlore antiferromagnet.  In particular, our microscopic calculation goes beyond the phenomenological approaches adopted
in most studies of magnetic excitations in helimagnets. Based on a well established microscopic model
for the helical order, we exactly diagonalize the linearized Landau-Lifshitz-Gilbert (LLG) equation with a
large unit cell; the size of the unit cell determines the ordering wavevector of the magnetic spiral in the commensurate limit.
By comparing results obtained from different helical pitchs, we find a spin-wave spectrum that is characteristic of
helimagnons. Our work thus provides a solid example of helimagnons in frustrated pyrochlore lattice.
We also performed dynamical LLG simulations on large finite lattices and found an
overall agreement with the experimental data on CdCr$_2$O$_4$.

{\em Model Hamiltonian.}
Our starting point is a classical spin Hamiltonian that includes nearest-neighbor (NN) exchange interactions
and spin-orbit coupling manifested as the Dzyaloshinskii-Moriya (DM) interaction~\cite{dzyaloshinskii64,moriya60}
between NN spins:
\begin{eqnarray}
	\mathcal{H} = \sum_{\langle ij \rangle}
	\Bigl[\left(J + K_{ij}\right)\mathbf S_i\cdot\mathbf S_j
	+ \mathbf D_{ij}\cdot\left(\mathbf S_i\times\mathbf S_j\right)\Bigr].
\end{eqnarray}
The isotropic NN exchange can be recast into $(J/2)\sum_{\mathbf r} \left|\mathbf M(\mathbf r)\right|^2$
up to a constant, where $\mathbf M(\mathbf r)$ denotes the vector sum of spins on a tetrahedron at $\mathbf r$.
Minimization of this term requires the vanishing of total magnetic moment on every tetrahedron but
leaves an extensively degenerate manifold.
The lattice distortion introduces the exchange anisotropy $K_{ij}$ through 
magnetoelastic coupling. Here we consider tetragonal lattice distortions that preserve
the lattice translational symmetry. In particular, the odd-parity distortion breaks the inversion symmetry
and stabilizes a collinear spin order with wavevector $\mathbf q = 2\pi(0, 0, 1)$~\cite{tcher02,chern09}.
The DM term is allowed on the ideal pyrochlore lattice, where the bonds are not centrosymmetric as
required by the so-called Moriya rules~\cite{moriya60}. Moreover, the high symmetry of the pyrochlore lattice
completely determines the orientations of the DM vectors up to a multiplicative factor.
The explicit form of the vectors $\mathbf D_{ij}$ can be found in Ref.~\cite{elhajal05,chern10}.

To understand how the coplanar helical order is stabilized, we briefly review the continuum
theory based on a gradient expansion of the order parameters~\cite{chern06}.
The antiferromagnetic order on the pyrochlore lattice is characterized by three staggered order
parameters $\mathbf L_i$ $(i = 1, 2,3)$.
For example, $\mathbf L_1 = (\mathbf S_0 + \mathbf S_1 - \mathbf S_2 - \mathbf S_3)/4$ describes
the staggered spins on two opposite $[011]$ and $[01\bar 1]$ bonds, here the subscript $\mu = 0\sim 3$ denotes the four
sublattices of the pyrochlore lattice (Fig.~\ref{fig:spiral}).
The order parameter for the collinear order is 
$\mathbf L_1(\mathbf r) = \hat\mathbf n_1\, e^{i\mathbf q\cdot\mathbf r}$,
and $\mathbf L_2 = \mathbf L_3 = 0$, where $\mathbf q = 2\pi(0,0,1)$, and $\hat\mathbf n_1$ is an arbitrary unit vector.

Assuming a slow variation of staggered magnetizations in the 
spiral, we parameterize the
order parameters as: $\mathbf L_i(\mathbf r) = e^{i\mathbf q\cdot\mathbf r}\phi_i(\mathbf r)\,\hat\mathbf n_i(\mathbf r)$,
where $\hat\mathbf n_i$ are three orthogonal unit vectors, $\phi_1 \approx 1 - \frac{1}{2}(\phi_2^2 + \phi_3^2)$,
and $\phi_2, \phi_3 \ll \phi_1$~\cite{chern06}. In the $J \to \infty$ limit, vanishing of 
total magnetization on tetrahedra
requires $\hat\mathbf n_3 \parallel \partial_y\hat\mathbf n_1$, and the $\phi$ fields can be analytically expressed
in terms of the director fields $\hat\mathbf n_i$. The effective energy functional in the gradient approximation
can be solely expressed in terms of the director fields $\hat\mathbf n_1$
and $\hat\mathbf n_2\parallel \partial_y\hat\mathbf n_1\times\hat\mathbf n_1$~\cite{chern06}:
\begin{eqnarray}
	\label{eq:E}
	& & \!\!\!\!\!\! \mathcal{F} = D \!\int \! d\mathbf r
	\left[\hat\mathbf n_1\!\cdot\!\bigl(\hat\mathbf a\times{\partial_x\hat\mathbf n_1}
	+\hat\mathbf b \times {\partial_y \hat\mathbf n_1}
	-\hat\mathbf c \times {\partial_z \hat\mathbf n_1}\bigr)\right] \\	
	& &\!\! + K_u\!\int\!d\mathbf r
	\Bigl[\left(\partial_x\hat\mathbf n_1\right)^2 \!+\! \left(\partial_y\hat\mathbf n_1\right)^2
	\!+\! 2\left(\partial_z\hat\mathbf n_1\right)^2
	\!-\! \left(\hat\mathbf n_2\cdot\partial_z\hat\mathbf n_1\right)^2\Bigr]. \nonumber
\end{eqnarray}
Here $K_u$ and $D$ denote the strengths of exchange anisotropy and DM interactions, respectively.
The form of the DM terms suggests spiral solutions with $\hat\mathbf n_1$ rotating about one of the principal
axes and staying in the plane perpendicular to it. These special solutions 
were first obtained in Ref.~\onlinecite{chern06}.
For example, the solution $\hat\mathbf n_1 = (\cos Q y, 0, \sin Q y)$
describes a helical order with coplanar spins lying in the $ac$ plane; see Fig.~\ref{fig:spiral}.
This spiral magnetic order produces a Bragg scattering at $\mathbf Q = 2\pi(0, \delta, 1)$,
consistent with the experimental characterization of CdCr$_2$O$_4$~\cite{chung05}.

The energy functional $\mathcal{F}$ actually possesses a continuous symmetry related to the O(2) rotational
invariance of the helical axis. To describe the general coplanar spiral solution,
we first introduce two orthogonal unit vectors $\hat\mathbf e_\xi$, $\hat\mathbf e_\zeta$
which lie in the 
$xy$ plane. A coplanar spiral rotating about the $\hat\mathbf e_\xi$ direction
has the form $\hat\mathbf n_1(\mathbf r) = \cos\theta(\xi)\,\hat\mathbf e_\zeta + \sin\theta(\xi)\,\hat\mathbf c$,
where $\xi \equiv \mathbf r\cdot\hat\mathbf e_\xi$. Substituting $\hat\mathbf n_1(\mathbf r)$ into Eq.~(\ref{eq:E})
yields an energy density:
$
	E[\theta] = -D \partial_{\xi}\theta + K_u \left(\partial_\xi\theta\right)^2/4.
$
Minimization of $E$ gives a helical wavenumber
\begin{eqnarray}
	Q = 2\pi\delta = d\theta/d\xi = 2D/K_u.
	\label{eq:delta}
\end{eqnarray}
This rotational symmetry is accidental and is lifted by further neighbor interactions and cubic symmetry field of the lattice.
A similar O(3) symmetry also occurs in the spiral order of MnSi~\cite{belitz06}.
In addition, all spirals whose axis lies in the $ab$ plane are also accidentally degenerate with the spiral in which spins rotate
about the $c$ axis~\cite{chern06}. The experimentally observed $(0,\delta,1)$ spiral is selected by a large third-neighbor
exchange $J_3$ which favors coplanar spirals rotating about the $a$ or $b$ axes.
The explicit on-site magnetizations of a spiral with the pitch vector parallel to the $b$-axis are
\begin{eqnarray}
	{\bar \mathbf S}_\mu(\mathbf r) = \hat\mathbf a\,\sin(\mathbf Q\cdot\mathbf r+\varphi_{\mu})
	+ \hat\mathbf c\,\cos(\mathbf Q\cdot\mathbf r+\varphi_{\mu}),
	\label{eq:S_bar}
\label{eq:Sm}
\end{eqnarray}
where $\mu = 0,1,2,3$ is the sublattice index, $\varphi_0=\psi$, $\varphi_1=\pi\delta + \psi$, $\varphi_2=\pi + \psi$,
and $\varphi_3=\pi(1+\delta) + \psi$, with $\psi$ is an arbitrary constant related to the U(1) symmetry of the coplanar spiral,
i.e. $\theta(\mathbf r) = \mathbf Q\cdot\mathbf r + \psi$.

\begin{figure*}
\centering{}
\includegraphics[width=1.7\columnwidth]{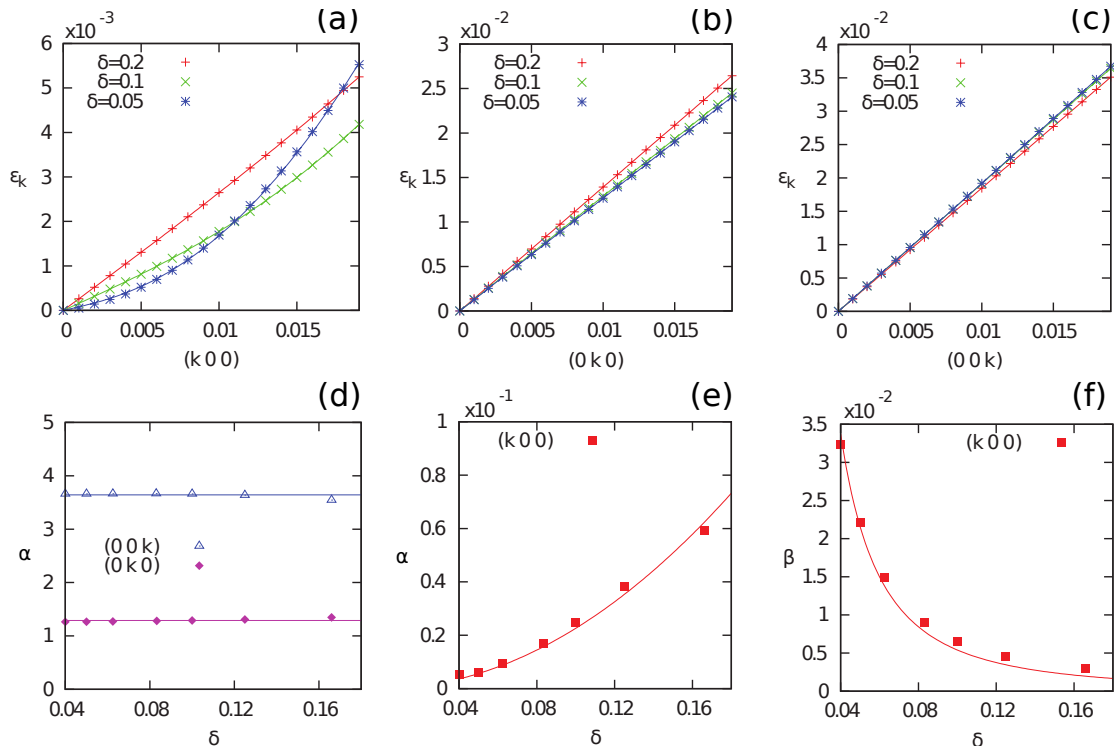}
\caption{(Color online) Spin-wave dispersions along three principal axes [panels (a)--(c)] obtained from numerical
exact diagonalization of a large unit cell. The results are fitted to the phenomenological
dispersion $\omega = \sqrt{\alpha k^2 + \beta k^4}$. The magnon energy is measured in unit of $JS$.
Panels~(d)--(f) show the dispersion parameters $\alpha$ and $\beta$ as a function of the 
helical wavenumber $\delta$. These data are fitted to $\alpha=$const, $\alpha \propto \delta^2$, 
and $\beta\propto \delta^{-2}$, respectively.
\label{fig:helimagnon}}
\end{figure*}

{\em Helimagnons in the continuum limit}.
We next consider a phenomenological approach to low-energy magnetic excitations in the helical ground state.
For a spiral with a fixed axis the obvious soft modes are the U(1) phase fluctuations $\psi$ associated with the
coplanar director field. A naive guess of the effective energy functional of the soft mode would
be $\mathcal{F} \propto K_u \int d\mathbf r |\nabla\psi|^2$, similar to that of planar magnets.
However, a phase variation $\psi = \alpha \zeta$ corresponding to a rotation of the helical axis is a
zero mode, a manifestation of the U(1) symmetry. Yet, it costs a finite amount of energy $\propto K_u \alpha^2$. 
Consequently, there cannot be any $(\partial_\zeta\psi)^2$ term in the effective energy density.
The correct energy functional for the soft modes is~\cite{deGennes93,chaikin,belitz06}
\begin{eqnarray}
	\label{eq:F}
	\mathcal{F} = \frac{1}{2}\!\int\! d\mathbf r \Bigl[c_z\!\left(\partial_z \psi\right)^2
	+\!c_{\parallel}\! \left(\partial_{\xi}\psi\right)^2
	+\! \frac{c_{\perp}}{Q^2}\! \left(\partial^2_{\zeta} \psi\right)^2 \! + r m^2 \Bigr], \,\,
\end{eqnarray}
where $c_\parallel$, $c_{\perp} \sim K_u$, $c_z \sim J_3$ are elastic constants,
and the true soft mode is a generalized phase accompanied by a rotation of the director field.
We have also included a zero wavevector ferromagnetic mode $m$ which is soft due to spin conservation.
It is worth noting that the first term in Eq.~(\ref{eq:F}) is absent in conventional helical order discussed in Ref.~\cite{belitz06}.
As discussed above, it originates from a rather large $J_3$ that gives penalty to variations along the $z$ direction.
Schematically the coarse-grained sublattice magnetizations are
\begin{eqnarray}
	\label{eq:Sr}
	& &\mathbf S_\mu(\mathbf r) \sim {\bar \mathbf S}_\mu(\mathbf r) \, + \, m(\mathbf r) \hat\mathbf b \\
	& & \qquad + \,\,\psi(\mathbf r)\,\bigl[\cos (\mathbf Q\cdot\mathbf r + \varphi_\mu)\,
	\hat\mathbf a - \sin (\mathbf Q\cdot\mathbf r + \varphi_\mu)\, \hat\mathbf c\bigr], \nonumber
\end{eqnarray}
The spin dynamics is governed by a generalized Landau-Lifshitz equation~\cite{chaikin}:
\begin{eqnarray}
	\frac{\partial \mathbf S_\mu(\mathbf r)}{\partial t} = -\gamma\, \mathbf S_\mu(\mathbf r) \times
	\frac{\delta \mathcal{F}}{\delta \mathbf S_\mu(\mathbf r)},
	\label{eq:LL-E}
\end{eqnarray}
where $\gamma$ is a constant. Substituting $\mathbf S_\mu$ and 
$\mathcal{F}$
into the Eq.~(\ref{eq:LL-E})
yields coupled differential equations $\partial_t \psi = \gamma r m$
and $\partial_t m = -\gamma (-c_{\parallel} \partial^2_\xi - c_z\partial_z^2 + c_{\perp}\partial^4_{\zeta}/Q^2)\psi$.
Assuming a space-time dependence $\exp(i\mathbf k\cdot\mathbf r - i \omega t)$,
we obtain a highly anisotropic dispersion relation~\cite{belitz06}
\begin{eqnarray}
	\omega(\mathbf k) = \gamma \, r^{1/2}\, \sqrt{c^{\phantom{1}}_z k_z^2
	+ c^{\phantom{1}}_{\parallel} k_{\parallel}^2
	+ c^{\phantom{1}}_{\perp} k_{\perp}^4 / Q^2}.
	\label{eq:omega}
\end{eqnarray}
For wavevectors parallel to the spiral and the $c$ axes the dispersion is linear
as in an antiferromagnet, while it is quadratic for wavevectors parallel to the $a$ axis.

{\em Exact diagonalization}.
At the microscopic level, the crystal symmetry anisotropy is expected to modify
the helimagnon dispersion~(\ref{eq:omega}). In addition, the particular spiral $(0,\delta,1)$ is stabilized
by a large $J_3$ which would increase the spin stiffness along the $c$ axis~\cite{chern06}. To take into account these
effects, here we perform exact diagonalizations with a large unit cell to investigate the
low-energy acoustic-like magnons and compare the results with predictions based on the helimagnon theory.

To begin with, we first choose a commensurate unit cell of 
a size which contains $2 \times \Lambda$ tetrahedra,
corresponding to an ordering wavevector $\mathbf q = 2\pi(0, \delta, 1)$ with $\delta = 1/\Lambda$.
Here the factor of 2 accounts for the staggering of spins along the $c$ axis. 
The DM constant is then determined
according to the expression $D = (3/\sqrt{2}) K_u\tan\pi\delta$ [the discrete version of Eq.~(\ref{eq:delta})] 
for a given fixed $K_u$.
Next we introduce small perturbations to spins in the ground state: $\mathbf S_i = S\,\hat{\bm \nu}_i + \bm\sigma_i$,
where~$\hat{\bm\nu}_i$ is a unit vector parallel to the magnetic moment at site $\mathbf r_i$ in the ground state,
and $\bm\sigma_i \perp \hat{\bm\nu}_i$ denotes transverse deviations. Note that because of this constraint,
there are only two degrees of freedom associated with $\bm\sigma_i$ at each site.
Assuming a time-dependence $\bm\sigma_i \sim \exp({i\mathbf k\cdot\mathbf r_i-i\omega t})$ 
for the eigenmodes, the linearized LLG equation can be cast into an eigenvalue
problem $\mathbb{T}(\mathbf k)\,\vec{\bm\sigma} = -i\omega_{\mathbf k}\, \vec{\bm\sigma}$,
where $\vec{\bm\sigma} = \left(\sigma_1, \sigma_2, \cdots, \sigma_n\right)$ is
a vector of dimension $n = 2\times 2\Lambda \times 4$, and $\mathbb T$ is an $n\times n$ matrix.
We then numerically diagonalize the matrix $\mathbb{T}({\bf k})$ and obtain the lowest eigenvalues in the vicinity
of ${\bf k} = 0$, corresponding to the soft-mode around the ordering vector of the magnetic spiral.

The magnon dispersions along the three principal directions in the vicinity of the ordering
wavevector $\mathbf Q=2\pi(0,\delta,1)$ are shown in Fig.~\ref{fig:helimagnon}(a)--(c).
The number of spins in the unit cell corresponding to used spiral pitch $\delta$ is $N_s = 40$, 80, 160.
To incorporate the anisotropy due to the cubic symmetry of the lattice, the obtained spectra
are fitted to the phenomenological dispersion: $\omega(k) = \sqrt{\alpha k^2 + \beta k^4}$.
The dispersion along the $b$ and $c$ axes can be well approximated by a linear relation, 
while the spectrum along the $a$-axis exhibits a predominant
quadratic behavior at small $\delta$ similar to that of a ferromagnet.
The dispersion parameters $\alpha$ and $\beta$ obtained from the fitting are shown in Fig.~\ref{fig:helimagnon}(d)--(f)
as a function of the helical wavenumber $\delta$. The large spin-wave velocity along the $c$ axis can be attributed
to a large $J_3$, consistent with the analysis of gradient expansion~\cite{chern06}.
More importantly, the $\delta$-dependence of parameters for dispersion along the $a$-axis exhibits
a scaling relation $\alpha \sim \delta^2$ and $\beta \sim 1/\delta^2$, reminiscent of that of a helimagnon~\cite{belitz06}.

\begin{figure}
\begin{centering}
\includegraphics[width=0.68\columnwidth]{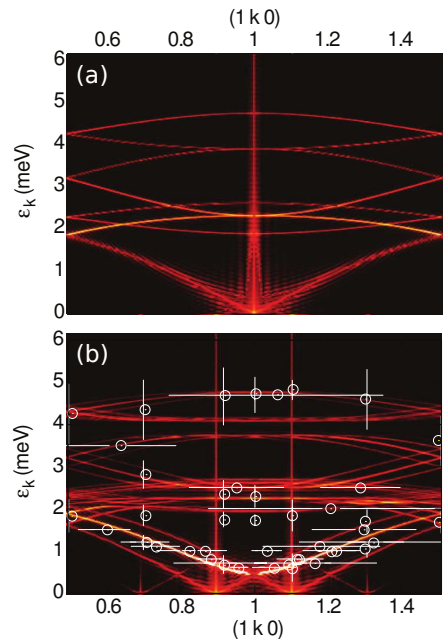}
\par\end{centering}
\caption{(Color online) Simulated spectra of (a) N\'eel state with $\mathbf q = 2\pi(0,0,1)$
and (b) helical order with $\mathbf q$ shifted to $2\pi(0,\delta,1)$ where the helical pitch $\delta=0.1$ 
due to a finite $D = 0.14$ meV.
Other parameters used in the simulation are $J=1.35$ meV, $K_{u}=0.21$ meV, and $J_{3}=0.28$ meV.
(b)~Fitted spin wave spectrum to the neutron scattering measurement (white circles)~\cite{chung05}.
The finite DM interaction splits each band into three as shown
in panel~(b). The vertical peaks at the ordering wavevectors are numerical artifacts.
\label{fig:fitting}}
\end{figure}

{\em Magnons in CdCr$_2$O$_4$}.
Although the exact diagonalization method discussed above provides a rather accurate magnon spectrum,
the complicated band-folding due to large unit cell makes it difficult to compare the results with experiment.
Instead, here we employ a less accurate but more direct approach based on simulating the real-time LLG
equations in a large finite lattice~\cite{mochizuki10}:
\begin{eqnarray}
	\label{eq:LLG}
	\frac{\partial\mathbf{S}_{i}}{\partial t}= \mathbf{S}_{i}\times
	\frac{\!\partial \mathcal{H}}{\partial \mathbf S_i}  +
	\frac{\alpha_G}{S}\mathbf{S}_{i}\times\frac{\partial\mathbf{S}_{i}}{\partial t},
\end{eqnarray}
where 
$\alpha_G$ is the dimensionless Gilbert-damping coefficient.
The dynamics simulation is initiated by a short pulse localized at $\mathbf r_i = 0$.
Since the system is driven by a white source with flat spectrum, we expect magnetic excitations
of various energy and momentum are generated in our simulations. The spin-wave spectrum is then obtained
via the Fourier transform of the simulation data.

The magnon spectrum  of $\mathrm{CdCr_{2}O_{4}}$ has been measured using inelastic neutron scattering
in Ref.~\cite{chung05}. Here we present our numerical calculations
and compare the results with the experimental findings.
The measured helical pitch $\delta \approx 0.1$ can be used to fix the ratio of the DM interaction
to the exchange anisotropy $K_u$. Then the absolute value of $K_u$ can be estimated by the magnon
energies at the zone center.
As for further-neighbor exchanges, {\it ab initio} calculations find a negligible $J_2$ and a quite large $J_3$
with antiferromagnetic sign~\cite{chern06,yaresko08}.
Using the following set of parameters: $J=1.35$ meV, $K_{u}=0.21$ meV, $D=0.14$ meV, and $J_{3}=0.28$ meV,
we obtained very good agreement with the experimentally measured spectrum, as shown in Fig.~\ref{fig:fitting}(b).

{\em Concluding discussions}.
We have studied the spin-wave excitations in a complex helical magnetic order on the pyrochlore lattice.
Our exact diagonalization calculation of the low-energy dispersion in the commensurate limit
of the long-period spiral clearly demonstrates the existence of helimagnon excitations in the frustrated
pyrochlore antiferromagnet. This novel magnetic excitation can be observed in spinel CdCr$_2$O$_4$
and itinerant magnet YMn$_2$.
We have performed dynamical LLG simulations with parameters inferred from structural
data and {\em ab initio} calculations. The numerical magnon spectrum agrees well with the measured dispersion
in CdCr$_2$O$_4$. Our work thus helps clarify the underlying mechanisms
that stabilize the observed helical order and provides a quantitative description of the model Hamiltonian.

{\em Acknowledgements}.
We thank J. Deisenhofer, O. Sushkov, O. Tchernyshyov for stimulating discussions.
N.P. and E.C. acknowledge the support from NSF grant DMR-1005932.
G.W.C. is supported by ICAM and NSF Grant DMR-0844115.
N.P. and G.W.C. also thank the hospitality of the visitors program at MPIPKS,
where part of the work on this manuscript has been done.

\end{document}